\documentclass[a4paper,12pt]{article}
\usepackage[pctex32]{graphics}
\usepackage{amssymb,amsmath}

\begin{document}

\topmargin 0pt
\oddsidemargin 0mm
\newcommand{\be}{\begin{equation}}
\newcommand{\ee}{\end{equation}}
\newcommand{\ba}{\begin{eqnarray}}
\newcommand{\ea}{\end{eqnarray}}
\newcommand{\fr}{\frac}

\renewcommand{\thefootnote}{\fnsymbol{footnote}}

\begin{titlepage}

\vspace{5mm}
\begin{center}
{\Large \bf Quasinormal modes for topologically massive black
hole}

\vspace{12mm}

{\large   Hyung Won Lee,
 Yong-Wan Kim,
and Yun Soo Myung\footnote{e-mail
 address: ysmyung@inje.ac.kr}}
 \\
\vspace{10mm}

{\em Institute of Mathematical Science and School of Computer
Aided Science, \\Inje University, Gimhae 621-749, Korea}
\end{center}

\vspace{5mm}

\centerline{{\bf{Abstract}}}

\vspace{5mm}

We calculate quasinormal modes of a massive scalar perturbation on
the topologically massive black hole (CS-BTZ black hole). The
chiral point of $\mu \ell=1$ corresponds to a newly extremal black
hole. We show that there is no quasinormal modes at this point.
Accordingly, we prove the unitarity of the CS-BTZ black hole at
the chiral point.

\vspace{5mm}

\end{titlepage}

\newpage

\renewcommand{\thefootnote}{\arabic{footnote}}
\setcounter{footnote}{0} \setcounter{page}{2}

\section{Introduction}
The gravitational Chern-Simons terms in three-dimensional Einstein
gravity produces a physically propagating massive
graviton~\cite{DJT}. This theory with a negative cosmological
constant $\Lambda=-1/\ell^2$ gives us the BTZ solution with mass
$m$ and angular momentum $j$ as a trivial
solution~\cite{BTZ,BHTZ}. However, there exists also the mixed
solution like~\cite{KL,Sol}
\begin{equation} \label{mma}
M=m+\frac{j}{\mu \ell^2},~~J=j+\frac{m}{\mu}.
\end{equation}
Hence, the horizon radius was  shifted due to the presence of
Chern-Simons term and thus the CS-BTZ black hole appeared as a new
solution~\cite{MKL}. In this case, the authors  found a newly
extremal black hole at the chiral point of $\mu \ell=1$, in
addition to the extremal BTZ black hole at $\ell m=j$.

Recently, Sachs and Solodukhin have calculated  quasinormal modes
of a massive graviton on the non-rotating BTZ black hole
background with the correspondence of $\mu l \leftrightarrow
-m$~\cite{SSol}. They have showed that quasinormal modes are
absent for the case of $\mu \ell=1$ with the central charges
$c_{\rm L}=0$ and $c_{\rm R}=3\ell/G_3$ because of the zero
right-moving conformal weight $h_{\rm R}(\mu)|_{\mu=1/\ell}=0$.
This implies that the quasinormal mode of a massive graviton is
always zero, since the massive graviton becomes a massless
left-moving graviton, an unphysical propagation. However, we
remind the reader that two horizons of CS-BTZ black hole becomes
degenerate at the chiral point of $\mu \ell=1$~\cite{LSS1}. This
means that in order to see the effect of Chern-Simons term on the
black hole, one needs to calculate quasinormal modes of a massive
scalar on the CS-BTZ black hole background.

In this Letter, we compute quasinormal modes of a massive scalar
perturbation  on the CS-BTZ black hole. We show that there is no
quasinormal modes at the chiral point. Accordingly, we prove the
unitarity of the CS-BTZ black hole at the chiral point.

\section{Scalar modes}
We start with the topologically mass gravity in anti-de Sitter
spacetimes~\cite{DJT}
\begin{equation}
I_{\rm TMG}=\frac{1}{16 \pi G_3}\int
d^3x\sqrt{-g}\Bigg[R+\frac{2}{\ell^2}-\frac{1}{2\mu}\varepsilon^{\lambda\mu\nu}\Gamma^\rho_{~\lambda\sigma}
\Big(\partial_{\mu}\Gamma^\sigma_{~\nu\rho}+\frac{2}{3}\Gamma^\sigma_{~\mu\tau}\Gamma^\tau_{~\nu\rho}\Big)\Bigg],
\end{equation}
where $\varepsilon$ is the tensor density defined by
$\epsilon/\sqrt{-g}$ with $\epsilon^{012}=1$. The $1/\mu$-term is
the first higher derivative correction in three dimensions because
it is the third-order derivative.

Varying  the this action  leads to
\begin{equation}
G_{\mu\nu}+\frac{1}{\mu}C_{\mu\nu}=0,
\end{equation}
where the Einstein tensor including the cosmological constant is
given by
\begin{equation}
G_{\mu\nu}=R_{\mu\nu}-\frac{R}{2}g_{\mu\nu}
-\frac{1}{\ell^2}g_{\mu\nu}
\end{equation}
and the Cotton tensor is
\begin{equation}
C_{\mu\nu}= \varepsilon_\mu^{~\alpha\beta}\nabla_\alpha
\Big(R_{\beta\nu}-\frac{1}{4}g_{\beta\nu}R\Big).
\end{equation}
The  BTZ black hole solution  is given by
\begin{equation}
ds^2_{BTZ} = -f(r) dt^2 + \frac{dr^2}{f(r)} + r^2 \Big(d\phi +
N^\phi(r) dt\Big)^2 , \label{btzmetric}
\end{equation}
where the metric function $f$ and the lapse function $N^\phi$ are
\begin{eqnarray}
f(r) &=& -8 G_3 m + \frac{r^2}{\ell^2} + \frac{16 G_3^2 j^2}{r^2} = \frac{(r^2 - r_+^2)(r^2 - r_-^2)}{\ell^2 r^2},
 \label{def_f} \\
N^\phi(r) &=& - \frac{4 G_3 J}{r^2} = - \frac{r_+ r_-}{\ell r^2}
\label{def_N}
\end{eqnarray}
with two horizons
\begin{equation}
r_{\pm} = \ell \left [ \sqrt{2 G_3 \left ( m +
\frac{j}{\ell}\right )}
             \pm \sqrt{2 G_3 \left ( m - \frac{j}{\ell}\right )} \right ]. \label{def_hor}
 \end{equation}
 Here $m$ and $j$ are the mass and angular momentum of the BTZ
 black hole, respectively.

On the other hand, plugging Eq.(\ref{mma}) into
Eq.(\ref{btzmetric}) ($M\to m,J\to j$) leads to the CS-BTZ black
hole solution~\cite{MKL} as
\begin{equation}
ds^2_{CS-BTZ} = -\tilde{f}(r) dt^2 + \frac{dr^2}{\tilde{f}(r)} +
r^2 (d\phi + \tilde{N}^\phi(r) dt)^2 , \label{btzmetrict}
\end{equation}
where
\begin{eqnarray}
\tilde{f}(r) &=& -8 G_3 \left (m+ \frac{j}{\mu \ell^2} \right ) +
\frac{r^2}{\ell^2} +
      \frac{16 G_3^2 (j+m/\mu)^2}{r^2} = \frac{(r^2 - \tilde r_+^2)(r^2 - \tilde r_-^2)}{\ell^2 r^2} \label{def_f_t} \\
\tilde{N}^\phi(r) &=& - \frac{4 G_3 (j+m/\mu)}{r^2} = -
\frac{\tilde r_+ \tilde r_-}{\ell r^2}. \label{def_N_t}
\end{eqnarray}
Here we have two shifted horizons
\begin{equation}
\tilde r_{\pm} = \ell \left ( \sqrt{ 1 + \frac{1}{\mu \ell}}
\sqrt{2 G_3 \left ( m + \frac{j}{\ell}\right )}
             \pm \sqrt{ 1 - \frac{1}{\mu \ell}} \sqrt{2 G_3 \left ( m - \frac{j}{\ell}\right )} \right ).
              \label{def_hor_t}
\end{equation}
We observe that the degenerate horizon of
$\tilde{r}_+=\tilde{r}_-\equiv\tilde{r}_{\rm e}$ appears at $\mu
\ell=1$ (chiral point) and $ \ell m=j$. The former one is a newly
extremal black hole due to the Chern-Simons term and the latter
exists even for the absence of the Chern-Simons term.

In order to obtain quasinormal modes~\cite{BSS}, we introduce the
wave equation for a massive  scalar field $\Psi$ on the CS-BTZ
black hole background of Eq.(\ref{btzmetrict})
\begin{equation}
(\nabla^2_{\rm CS-BTZ} - m^2) \Psi(t, r, \phi) = 0, \label{eom}
\end{equation}
where $m$ is the mass of perturbed field with $m^2 \ge 0$.
Considering the axially symmetric background, we parameterize the
perturbed field as
\begin{equation}
\Psi(t, r, \phi) = e^{-i \omega t} e^{i n \phi} \psi(r)
\label{wave_func}
\end{equation}
with angular number $n \in Z$. Then, the wave  equation can be
written as
\begin{equation}
\tilde{f} \frac{d^2 \psi(r)}{dr^2} + \left ( \frac{\tilde{f}}{r} +
\frac{d\tilde{f}}{dr} \right ) \frac{d\psi(r)}{dr} + \left(
\frac{(\omega + \tilde{N}^\phi n )^2}{\tilde{f} } -
\frac{n^2}{r^2} - m^2\right ) \psi(r) = 0. \label{r_eq}
\end{equation}
Introducing  a new variable $z = (r^2 - \tilde r_+^2)/(r^2 -
\tilde r_-^2)$, the above  equation can be  transformed into
\begin{equation}
z(1-z) \frac{d^2 \psi(z)}{dz^2} + (1-z) \frac{d\psi(z)}{dz} +
\left (  \frac{A}{z} + B + \frac{C}{1-z} \right ) \psi(z) = 0,
\label{zeq}
\end{equation}
where
\begin{eqnarray}
A &=& \frac{\ell^4}{4(\tilde{r}_+^2 - \tilde{r}_-^2)^2}
 \left ( \omega \tilde r_+ - \frac{n}{\ell} \tilde r_- \right )^2 \equiv \alpha_+^2, \label{def_A} \\
B &=& - \frac{\ell^4}{4(\tilde{r}_+^2 - \tilde{r}_-^2)^2} \left ( \omega \tilde r_- - \frac{n}{\ell}
 \tilde r_+ \right )^2 \equiv - \alpha_-^2, \label{def_B} \\
C &=& -\frac{m^2 \ell^2}{4}. \label{def_C}
\end{eqnarray}
 From now on, we assume $\alpha_{\pm} > 0$ to have a propagating wave.
The general solution to Eq.(\ref{zeq}) at the near horizon  of  $z
\sim 0$  is given by the hypergeometric function ${}_2F_1$  with
two constants $C_1$ and $C_2$ as
\begin{equation}
\psi(z) = z^{i \alpha_+} (1-z)^{(1-\Delta)/2} \left [ C_1 \, {}_2F_1(a, b;c;z) + C_2 z^{1-c} {}_2F_1(a+1-c, b+1-c;2-c;z) \right ],
\label{zero_sol}
\end{equation}
where the parameters $\Delta,~a,~ b,$ and $ c$ take the forms
\begin{eqnarray}
\Delta &=& \sqrt{1+m^2 \ell^2}, \label{def_Delta} \\
a &=& \frac{1}{2}(1 - \Delta) + i(\alpha_+ + \alpha_-), \label{def_a} \\
b &=& \frac{1}{2}(1 - \Delta) + i(\alpha_+ - \alpha_-), \label{def_b} \\
c &=& 1 + 2i\alpha_+. \label{def_c}
\end{eqnarray}
We note  the solution of the limiting case ($z \rightarrow 0$)
\begin{equation}
\psi_{z \rightarrow 0}(z) = C_1 z^{i \alpha_+} + C_2 z^{-i
\alpha_+}. \label{zero_limit}
\end{equation}
Then, we obtain the corresponding solution $\Psi(t, r, \phi)$ as
\begin{equation}
\Psi_{z \rightarrow 0}(t,z,\phi) = e^{in \phi} \left \{ C_1
e^{-i(\omega t - \alpha_+ \ln z)} + C_2 e^{-i(\omega t + \alpha_+
\ln z)} \right \}, \label{zero_limit_f}
\end{equation}
where the first term corresponds to an outgoing wave, while  the
second is  an ingoing wave at the outer horizon. We note that the
boundary condition for the quasinormal modes in  asymptotically
AdS spacetimes: ingoing mode (ingoing flux) at the outer horizon
and no non-normalizable mode (zero flux) at infinity.

Since we have  an ingoing wave at the  horizon, it requires that
$C_1 = 0$. Hence the solution at the near horizon is given by
\begin{equation}
\psi(z) = C_2 z^{-i \alpha_+} (1-z)^{(1-\Delta)/2} \,
{}_2F_1(a+1-c, b+1-c;2-c;z). \label{sol_hor}
\end{equation}
Now we are in a position to  obtain the solution at infinity ($z =
1$), by applying the transformation formula for hypergeometric
functions. The result is
\begin{eqnarray}
\psi(z) &=& C z^{-i \alpha_+} (1-z)^{(1-\Delta)/2} \times \nonumber \\
&&\left \{
\frac{\Gamma(2-c) \Gamma(c-a-b-1)}{\Gamma(1-a)\Gamma(1-b)} \, {}_2F_1(a+1-c, b+1-c;a+b+1-c;1-z) \right . \nonumber \\
&+&\left . (1-z)^{c-a-b}\frac{\Gamma(2-c)
\Gamma(a+b-c)}{\Gamma(a-c+1)\Gamma(b-c+1)} \, {}_2F_1(1-a,
1-b;c-a-b+1;1-z) \right \}, \label{sol_inf1}
\end{eqnarray}
where the first term corresponds to the non-normalizable mode and
the second is the normalizable one. In order to obtain the
normalizable term at infinity only, we  require the conditions
\begin{equation}
1-a = -k, \hspace{0.5cm} {\rm or} \hspace{0.5cm} 1-b = -k
\label{bc}
\end{equation}
with mode number $k = 0, 1, 2, \cdots$. The overtones are defined
as $k\not=0$ if the quasinormal modes are present.  From this
condition, we find the two types of quasinormal modes for
$\tilde{r}_+ \not=\tilde{r}_-$ as
\begin{eqnarray}
\omega_{\rm L} &=& \frac{n}{\ell} -2i \left ( \frac{\tilde r_+ - \tilde r_-}{\ell^2} \right )
           \left ( k + \frac{1}{2} + \frac{\sqrt{1+m^2\ell^2}}{2}\right ), \label{left_w} \\
\omega_{\rm R} &=& -\frac{n}{\ell} -2i \left ( \frac{\tilde r_+ + \tilde r_-}{\ell^2} \right )
           \left ( k + \frac{1}{2} + \frac{\sqrt{1+m^2\ell^2}}{2}\right )
           \label{right_w},
\end{eqnarray}
where $\omega_{\rm L}$ describes the left-moving degrees of
freedom and $\omega_{\rm R}$ describes the right-moving degrees of
freedom, in accordance with the CFT picture on the boundary at
infinity~\cite{BSS}. Expressing these as the parameters of the
CS-BTZ black holes leads to
\begin{eqnarray}
\omega_{\rm L} &=& \frac{n}{\ell} -\frac{4i}{\ell}
\sqrt{1-\frac{1}{\mu\ell}} \sqrt{2 G_3 \left( m -
\frac{j}{\ell}\right)}
           \left ( k + \frac{1}{2} + \frac{\sqrt{1+m^2\ell^2}}{2}\right ) \label{leftwm} \\
\omega_{\rm R} &=& -\frac{n}{\ell} -\frac{4i}{\ell}
\sqrt{1+\frac{1}{\mu\ell}} \sqrt{2 G_3 \left( m +
\frac{j}{\ell}\right)}
           \left ( k + \frac{1}{2} + \frac{\sqrt{1+m^2\ell^2}}{2}\right ) \label{rightwm}.
\end{eqnarray}
The normalizable solution at infinity is then  given by
\begin{equation}
\psi_{z\to 1}(z) = C_2 z^{-i \alpha_+} (1-z)^{(1+\Delta)/2}
\frac{\Gamma(2-c) \Gamma(a+b-c)}{\Gamma(a-c+1)\Gamma(b-c+1)} \,
{}_2F_1(1-a, 1-b;c-a-b+1;1-z). \label{sol_inf}
\end{equation}

We observe what happens as $\mu$ approaches  the chiral point
($\mu \ell \to 1$). In this case, apparently,  we have
$\omega_{\rm L}\to n/\ell - i0$ while $\omega_{\rm R} \to
-n/\ell-i4(\tilde{r}_{\rm e}/\ell^2)(k+1/2+\Delta/2)$. This may be
consistent with the picture of chiral gravity with central charges
$c_{\rm L}=0$ and $c_{\rm R}=3l/G_3$. However, to obtain
quasinormal modes with $ \omega_{\rm R}$ on the extremal black
hole background, we have to make a further study because the
coordinate of $z$ is not appropriate for describing the extremal
case of $\tilde{r}_+=\tilde{r}_-$. Unfortunately, we have the same
point of $z=1$ at infinity of $r\to \infty$ and the extremal point
of $\tilde{r}_+=\tilde{r}_-$.

\section{Scalar modes at the chiral point}
For the extremal black holes, we have the horizon radius as
\begin{equation}
\tilde r_{\rm e} = \ell \sqrt{ 1 + \frac{1}{\mu \ell}} \sqrt{2 G_3
\left ( m + \frac{j}{\ell}\right )}. \label{extremal_hor}
\end{equation}
For $\mu \ell=1$, we have $\tilde{r}_{\rm
e}=2\ell\sqrt{G_3(m+j/\ell)}$, while for $m \ell=j$, we have
$\tilde{r}_{\rm e}=2\ell\sqrt{1+1/\mu \ell}\sqrt{G_3m}$. In this
section, we consider the first case only.  To solve the wave
equation (\ref{r_eq}) on the extremal CS-BTZ black hole
background~\cite{CLS,ML}, we introduce a new variable $\tilde z$
as
\begin{equation}
\tilde z = - 2 i \Omega_- \frac{\tilde{r}_{\rm e}^2}{r^2 - \tilde
r_{\rm e}^2}, \label{tildez_def}
\end{equation}
where
\begin{equation}
\Omega_{\pm} = \frac{\ell^2}{2 \tilde {r}_{\rm e}} \left (
\omega_{\rm e} \pm \frac{n}{\ell} \right ). \label{Omega_def}
\end{equation}
Then, Eq.(\ref{r_eq}) becomes
\begin{equation}
\frac{d^2 \psi}{d \tilde z^2} + \left \{ -\frac{1}{4} + \frac{i
\Omega_+}{2} \frac{1}{\tilde z} - \frac{m^2 \ell^2}{4}
\frac{1}{\tilde z^2}\right \} \psi =0, \label{tildez_eq}
\end{equation}
which is the Whittaker's equation~\cite{Abr70}. Its solution is
given by
\begin{equation}
\psi(\tilde z) = \tilde{C}_1 M\Big(\frac{i}{2}\Omega_+,
\frac{\Delta}{2}, \tilde z\Big) +
                 \tilde{C}_2 W\Big(\frac{i}{2}\Omega_+, \frac{\Delta}{2}, \tilde z\Big) ,
\label{tildez_sol}
\end{equation}
where $M$ is the Whittaker M-function and  $W$ is the Whittaker
W-function. Here  $\Delta = \sqrt{1+m^2 \ell^2}$. They are related
to confluent hypergeometric functions $F$ and $U$, respectively
\begin{eqnarray}
M(a, b, \tilde z) = e^{-\tilde z/2} {\tilde z}^{1/2+b} F(1/2+b-a, 1+2b, \tilde z), \label{rel_mf} \\
W(a, b, \tilde z) = e^{-\tilde z/2} {\tilde z}^{1/2+b} U(1/2+b-a,
1+2b, \tilde z). \label{rel_wu}
\end{eqnarray}
For $\tilde z \rightarrow 0$, $W \sim \tilde
z^{\frac{1-\Delta}{2}}$ diverges. Thus, we choose $\tilde{C}_2 =
0$ to have a normalizable mode at infinity. We obtain a final
solution at infinity
\begin{equation}
\psi(\tilde z) = \tilde{C}_1 M\Big(\frac{i}{2}\Omega_+,
\frac{\Delta}{2}, \tilde z\Big). \label{extreme_sol_inf}
\end{equation}
Now we can derive an explicit solution  near the horizon at
$\tilde z = \infty (r = \tilde r_e)$. For this purpose, we
introduce an  asymptotic expansion of the confluent hypergeometric
function
\begin{equation} \label{asymp_expansion}
F(a,c;\tilde z) \to
\frac{\Gamma(c)}{\Gamma(c-a)} \tilde z^{-a} e^{ \pm i\pi a}+
\frac{\Gamma(c)}{\Gamma(a)}\tilde z^{a-c} ,
\end{equation}
where the upper sign being taken if $-\pi/2 <{\rm arg} (\tilde z)
< 3\pi/2$ and the lower one if $-3\pi/2 <{\rm arg} (\tilde z) \le
-\pi/2$. Using this relation, we find the solution  near the
horizon as
\begin{eqnarray}
\psi(\tilde z) &\simeq& e^{-\tilde z/2} \tilde z^{-(1+\Delta)/2+i\Omega_+/2} \frac{\Gamma(1+\Delta)}{\Gamma(1/2+\Delta/2+i\Omega_+/2)}
                    e^{\pm i \pi(1/2+\Delta/2-i\Omega_+/2)} \nonumber \\
&+& e^{\tilde z /2} \tilde z^{(1+\Delta)/2-i\Omega_+/2} \frac{\Gamma(1+\Delta)}{\Gamma(1/2+\Delta/2-i\Omega_+/2)} \nonumber \\
&\equiv& \psi^{\rm in} (\tilde z)+ \psi^{\rm out}(\tilde z) .
\label{extreme_sol_hor}
\end{eqnarray}
In order to obtain quasinormal modes,  the wave function should be
a purely incoming mode near the event horizon. This requirement
could be accomplished  by demanding  $\psi^{\rm out}(\tilde{z})=0$
as
\begin{equation}
\frac{1}{2} + \frac{\Delta}{2} - \frac{i}{2} \Omega_+ = -k,
\label{quascond}
\end{equation}
for $k = 0, 1, 2, \cdots$. Inverting this expression leads to the
quasinormal frequencies
\begin{equation}
\tilde{\omega}_{\rm e} = -\frac{n}{\ell} - 4 i
\frac{\tilde{r}_{\rm e}}{\ell^2} \left ( k + \frac{1}{2} +
\frac{\sqrt{1+m^2\ell^2}}{2} \right ), \label{extreme_quasi}
\end{equation}
which is the same form recovered from  $\omega_{\rm R}$ in
Eq.(\ref{right_w}) with $\mu \ell=1$. However, the condition of
Eq.(\ref{quascond}) leads in turn to the zero ingoing flux. This
is because the flux expression
\begin{equation}
\label{extrema_flux_hor}
{\cal F}_{\rm in}(\tilde z=\infty) \propto
 \frac{\Gamma(1+\Delta)}{\Gamma(1/2+\Delta/2-i\Omega_+/2)}
 \frac{\Gamma(1+\Delta)}{\Gamma(1/2+\Delta/2+i\Omega_+/2)}
\end{equation}
leads to zero exactly when choosing $1/2+\Delta/2-i\Omega_+/2=-k$.
 This
implies that there is no room to accommodate quasinormal modes
with $\omega_{\rm R}=\tilde{\omega}_{\rm e}$ of a massive scalar
on the background of the extremal CS-BTZ black hole. The absence
of quasinormal modes
 is consistent with the picture of a stable event
horizon with thermodynamic properties $T_{H}=C_{J}=0,S_{BH}=\pi
\tilde{r}_{\rm e}/2G_3$ of the extremal CS-BTZ black hole. This is
so because the presence of quasinormal modes implies that a
massive scalar wave is losing its energy continuously into the
extremal event horizon.

Furthermore, the absence of $\omega_{\rm L}$ implies that the
extremal CS-BTZ black hole is  chiral, in compared with
$\omega_{\rm L/R}$ in Eqs.(\ref{leftwm}) and (\ref{rightwm}) for
the non-extremal CS-BTZ black holes.

\section{Discussions}
We study the wave equation for  a  massive scalar  in
three-dimensional CS-BTZ black hole spacetimes to understand the
chiral point of $\mu \ell=1$. Here we introduce two interesting
black hole spacetimes: the CS-BTZ and  extremal CS-BTZ.
 In the
CS-BTZ case, one finds  quasinormal modes. The presence of
quasinormal modes means that it shows a leakage of information
into the event horizon (dissipative object) and thus it  signals a
breakdown of the unitarity.

On the other hand, we do not find any quasinormal modes for the
extremal CS-BTZ black hole. Instead, we find a real and continuous
frequency $\omega_{\rm e}$ for the extremal CS-BTZ case~\cite{CLS}
which implies that the extremal CS-BTZ is a unitary system. We
show that the radial flux is identically zero outside the event
horizon, even though their wave functions are non-zero. Actually,
we obtain the ingoing flux as well as the outgoing flux, but
summing over these develops the zero flux near the event horizon
exactly. This means that there is no leakage of information into
the event horizon. Furthermore, the absorption cross section to
the extremal CS-BTZ may vanish~\cite{GM} and thus this extremal
black hole may be transparent for any wave energy.  Hence we argue
that the extremal CS-BTZ black hole is a unitary system~\cite{ML}.
In this case, we cannot obtain discrete spectra like $\omega_{\rm
e}=-n/\ell$ because this belongs to the non-compact system.

On the other hand, the linearized equation for the massive
graviton mode could not be represented by the massive scalar
equation (\ref{eom}). The reason is that the massive scalar could
represent the case of the same highest weights of $(h_{\rm
L},h_{\rm R}$) with $h_{\rm L}=h_{\rm R}$. However, three graviton
modes are massless left-moving, massless right-moving, and massive
modes.  They have the highest weights as (2,0), (0,2), and
$(3/2+\mu \ell/2,-1/2+\mu \ell/2)$, respectively. The massive
gravitons have negative energy for $\mu \ell >1$. At the critical
point of $\mu \ell=1$, its highest weights reduce to $(2,0)$,
those of a massless left-moving graviton. This corresponds to an
unphysically propagating gauge boson. Considering conformal
dimension $\Delta=h_{\rm L}+h_{\rm R}$, this boson may belong to
the massive scalar with $\Delta=2(m^2\ell^2=8)$. However, it has
non-zero spin of $s=h_{\rm L}-h_{\rm R}=2$. Hence, the quasinormal
modes of graviton modes (gauge bosons) with spin could not be read
off from the massive scalar modes with spin zero on the extremal
CS-BTZ background. We propose to use the spin-dependent wave
equation in Ref.\cite{KLM}, instead of Eq.(\ref{eom}).

\section*{Acknowledgement}
H.W. Lee was in part supported by KOSEF, Astrophysical Research
Center for the Structure and Evolution of the Cosmos at Sejong
University. Y. Myung  was in part supported by  the Korea Research
Foundation (KRF-2006-311-C00249) funded by the Korea Government
(MOEHRD).

\end{document}